\documentclass[twoside,twocolumn,9pt]{article}
\usepackage{extsizes}
\usepackage[super,sort&compress,comma]{natbib} 
\usepackage[version=3]{mhchem}
\usepackage[left=1.5cm, right=1.5cm, top=1.785cm, bottom=2.0cm]{geometry}
\usepackage{balance}
\usepackage{mathptmx}
\usepackage{sectsty}
\usepackage{graphicx} 
\usepackage{lastpage}
\usepackage[format=plain,justification=justified,singlelinecheck=false,font={stretch=1.125,small,sf},labelfont=bf,labelsep=space]{caption}
\usepackage{float}
\usepackage{fancyhdr}
\usepackage{fnpos}
\usepackage[english]{babel}
\addto{\captionsenglish}{%
  
}
\usepackage{array}
\usepackage{droidsans}
\usepackage{charter}
\usepackage[T1]{fontenc}
\usepackage[usenames,dvipsnames]{xcolor}
\usepackage{setspace}
\usepackage{upgreek}
\usepackage{siunitx}
\DeclareSIUnit\angstrom{\text {Å}}
\DeclareSIUnit\bohr{\text {\ensuremath {a}}_{0}}
\DeclareSIUnit\atom{\text atom}
 \DeclareSIUnit\bar{bar}
\usepackage{amsmath}
\usepackage{amsfonts}
\usepackage{amssymb}
\usepackage[compact]{titlesec}
\usepackage{hyperref}
\usepackage{booktabs}
\usepackage{caption}
\usepackage{subcaption}
\usepackage{tabularx}

\usepackage{subcaption}
\definecolor{cream}{RGB}{222,217,201}

\begin{document}

\pagestyle{fancy}
\thispagestyle{plain}

\newcommand{\alox}{$\upalpha$-Al$_2$O$_3$~}
\newcommand{\aloxp}{$\upalpha$-Al$_2$O$_3$}

%%%HEADER%%%
\fancypagestyle{plain}{
%%%HEADER%%%
\renewcommand{\headrulewidth}{0pt}
}
%%%END OF HEADER%%%

%%%PAGE SETUP%%% - Please do not change any commands within this section%%%
\makeFNbottom
\makeatletter
\renewcommand\LARGE{\@setfontsize\LARGE{15pt}{17}}
\renewcommand\Large{\@setfontsize\Large{12pt}{14}}
\renewcommand\large{\@setfontsize\large{10pt}{12}}
\renewcommand\footnotesize{\@setfontsize\footnotesize{7pt}{10}}
\makeatother

\renewcommand{\thefootnote}{\fnsymbol{footnote}}
\renewcommand\footnoterule{\vspace*{1pt}% 
\color{cream}\hrule width 3.5in height 0.4pt \color{black}\vspace*{5pt}} 
\setcounter{secnumdepth}{5}

\makeatletter 
\renewcommand\@biblabel[1]{#1}            
\renewcommand\@makefntext[1]% 
{\noindent\makebox[0pt][r]{\@thefnmark\,}#1}
\makeatother 
\renewcommand{\figurename}{\small{Fig.}~}
\sectionfont{\sffamily\Large}
\subsectionfont{\normalsize}
\subsubsectionfont{\bf}
\setstretch{1.125} %In particular, please do not alter this line.\setlength{\skip\footins}{0.8cm}
\setlength{\footnotesep}{0.25cm}
\setlength{\jot}{10pt}
\titlespacing*{\section}{0pt}{4pt}{4pt}
\titlespacing*{\subsection}{0pt}{15pt}{1pt}
%%%END OF PAGE SETUP%%%

%%%FOOTER%%%
\fancyfoot{}
\fancyfoot[LO,RE]{\vspace{-7.1pt}\includegraphics[height=9pt]{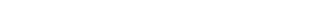}}
\fancyfoot[CO]{\vspace{-7.1pt}\hspace{13.2cm}\includegraphics{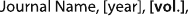}}
\fancyfoot[CE]{\vspace{-7.2pt}\hspace{-14.2cm}\includegraphics{head_foot/RF}}
\fancyfoot[RO]{\footnotesize{\sffamily{1--\pageref{LastPage} ~\textbar  \hspace{2pt}\thepage}}}
\fancyfoot[LE]{\footnotesize{\sffamily{\thepage~\textbar\hspace{3.45cm} 1--\pageref{LastPage}}}}
\fancyhead{}
\renewcommand{\headrulewidth}{0pt} 
\renewcommand{\footrulewidth}{0pt}
\setlength{\arrayrulewidth}{1pt}
\setlength{\columnsep}{6.5mm}
\setlength\bibsep{1pt}
%%%END OF FOOTER%%%

%%%FIGURE SETUP%%%- please do not change any commands within this section%%%
\makeatletter 
\newlength{\figrulesep} 
\setlength{\figrulesep}{0.5\textfloatsep} 
\newcommand{\topfigrule}{\vspace*{-1pt}%
\noindent{\color{cream}\rule[-\figrulesep]{\columnwidth}{1.5pt}} }
\newcommand{\botfigrule}{\vspace*{-2pt}%
\noindent{\color{cream}\rule[\figrulesep]{\columnwidth}{1.5pt}} }
\newcommand{\dblfigrule}{\vspace*{-1pt}%
\noindent{\color{cream}\rule[-\figrulesep]{\textwidth}{1.5pt}} }
\makeatother
%%%END OF FIGURE SETUP%%%

%%%TITLE, AUTHORS AND ABSTRACT%%%
\twocolumn[
    \begin{@twocolumnfalse}
{\includegraphics[height=30pt]{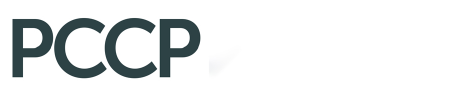}\hfill\raisebox{0pt}[0pt][0pt]{\includegraphics[height=55pt]{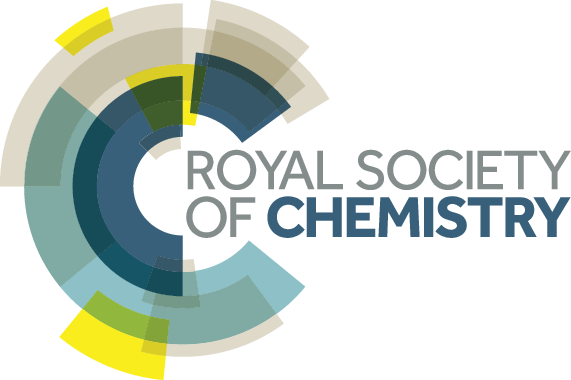}}\\[1ex]
\includegraphics[width=18.5cm]{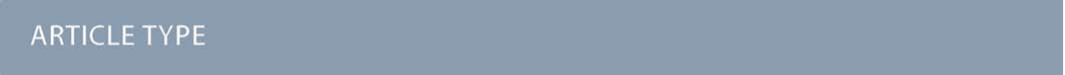}}\par
\vspace{1em}
\sffamily
\begin{tabular}{m{4.5cm} p{13.5cm} }

\includegraphics{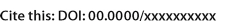} & 
\noindent\LARGE{\textbf{Hydrogen Atom Scattering at the Al$_2$O$_3$(0001) Surface: A Combined Experimental and Theoretical Study}} \\
\vspace{0.3cm} & \vspace{0.3cm} \\

 & \noindent\large{Martin Liebetrau$^{1,2}$, Yvonne Dorenkamp$^{3}$, Oliver B\"unermann$^{3,4,5,*}$ and J\"org Behler$^{1,2,**}$} \\

\includegraphics{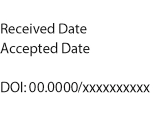} & \noindent\normalsize
{Investigating atom-surface interactions is the key to an in-depth understanding of chemical processes at interfaces, which are of central importance in many fields -- from heterogeneous catalysis to corrosion. 
In this work, we present a joint experimental and theoretical effort to gain insights into the atomistic details of hydrogen atom scattering at the \aloxp(0001) surface. Surprisingly, this system has been hardly studied to date, although hydrogen atoms as well as \alox are omnipresent in catalysis as reactive species and support oxide, respectively. We address this system
by performing hydrogen atom beam scattering experiments and molecular dynamics (MD) simulations based on a high-dimensional machine learning potential trained to density functional theory data. Using this combination of methods we are able to probe the properties of the multidimensional potential energy surface governing the scattering process. Specifically, we compare the angular distribution and the kinetic energy loss of the scattered atoms obtained in experiment with a large number of MD trajectories, which, moreover, allow to identify the underlying impact sites at the surface.
}

\end{tabular}

 \end{@twocolumnfalse} \vspace{0.6cm}
  ]
  
%%%END OF TITLE, AUTHORS AND ABSTRACT%%%
%%%FONT SETUP - please do not change any commands within this section
\renewcommand*\rmdefault{bch}\normalfont\upshape
\rmfamily
\section*{}
\vspace{-1cm}

%%%FOOTNOTES%%%

\footnotetext{\textit{$^{1}$}~Lehrstuhl f\"ur Theoretische Chemie II, Ruhr-Universit\"at Bochum, D-44780 Bochum, Germany}
\footnotetext{\textit{$^{2}$}~Research Center Chemical Sciences and Sustainability, Research Alliance Ruhr, D-44780 Bochum, Germany}
\footnotetext{\textit{$^{3}$}~Georg-August-Universit\"at G\"ottingen, Institut f\"ur Physikalische Chemie, Tammannstra{\ss}e 6, D-37077 G\"ottingen, Germany}
\footnotetext{\textit{$^{4}$}~Department of Dynamics at Surfaces, Max-Planck-Institute for Multidisciplinary Sciences, Am Fassberg 11, D-37007 Göttingen, Germany}
\footnotetext{\textit{$^{5}$}~International Center of Advanced Studies of Energy Conversion, Georg-August-Universit\"at G\"ottingen, Tammannstra{\ss}e 6, D-37077 G\"ottingen, Germany}
\footnotetext{\textit{$^{*}$}~Corresponding Author: oliver.buenermann@chemie.uni-goettingen.de}
\footnotetext{\textit{$^{**}$}~Corresponding Author: joerg.behler@rub.de}

%Please use \dag to cite the ESI in the main text of the article.
%If you article does not have ESI please remove the the \dag symbol from the title and the footnotetext below.
%\footnotetext{\dag~Electronic Supplementary Information (ESI) available: [details of any supplementary information available should be included here]. See DOI: 10.1039/b000000x/}

%%%END OF FOOTNOTES%%%

%%%MAIN TEXT%%%%

%%%%%%%%%%%%%%%%%%%%%%%%%%%%%%%%%%%%%%%%%%%%%%%%%%%
\section{Introduction}
%%%%%%%%%%%%%%%%%%%%%%%%%%%%%%%%%%%%%%%%%%%%%%%%%%%

Heterogeneously catalyzed reactions are of utmost economical importance, as a substantial part of the bulk products in chemical industry is made through reactions at solid catalyst surfaces \cite{doi:https://doi.org/10.1002/0470862106.ia084,FECHETE20122}. Prominent examples are the Haber-Bosch process for ammonia synthesis \cite{doi:10.1126/science.aar6611}, the cracking of long-chain hydrocarbons in the petroleum industry~\cite{doi:10.1021/acs.energyfuels.2c00567}, and the oxidation of sulphur dioxide to sulphur trioxide finally yielding sulphuric acid~\cite{https://doi.org/10.1002/zaac.202100091,JORGENSEN20074496}.
The improvement of such catalytic systems crucially depends on a detailed understanding of the reaction mechanisms at the atomic scale that are governed by the potential energy surface (PES), which is a high-dimensional function providing the energy and forces for any given atomic configuration. Thus, the PES determines the dynamics of the reactants at the surface, the making and breaking of bonds and, finally, product formation. 
\par
Insights into the interactions between atoms or molecules and surfaces can be obtained from both, experimental as well as theoretical studies. In particular atom and molecular beam experiments under well-defined conditions can provide invaluable details of processes at single-crystal surfaces\cite{Chadwick2017,Park2019,doi:10.1021/acs.jpca.1c00361}        . For instance, information about the interaction potential and energy transfer between atoms or molecules and surfaces can be derived from inelastic scattering experiments. Moreover, using rare-gas atoms, surface phonons of a material can be studied\cite{Benedek1994,Gumhalter2001}. Further, the role of rotational, vibrational and translational energy on surface reactivity is of central interest in molecule-surface scattering experiments\cite{Park2019,Werdecker2018,Alkoby2020,Kleyn1981,Greenwood2023}. Finally, scattering of open-shell atoms like hydrogen or oxygen provides detailed information about chemical bond formation in surface reactions \cite{doi:10.1021/acs.jpca.1c00361,Zhao2023}.
\par
Next to experiment, theoretical studies have become an indispensable tool in surface science and heterogeneous catalysis in the past two decades \cite{C8CS00398J,P2541,P1881,P0373}. Still, in spite of the substantial increase in computational power, first-principles methods like density functional theory (DFT), which have been very successfully applied to the calculation of properties like adsorption energies and surface structures, remain too demanding for determining a large number of molecular dynamics (MD) trajectories when computing the forces on-the-fly for each new MD step. Thus, it is common practice to perform extended MD simulations for scattering at surfaces employing atomistic potentials, which -- after fitting to accurate electronic structure data -- can provide a direct relation between atomic positions and forces while they are much cheaper to evaluate. 
A wide range of methods has been proposed to represent atomistic potentials for surface scattering processes \cite{P0221,P0449,P0220}, but constructing high-dimensional PESs of first-principles quality taking all the surface degrees of freedom explicitly into account has remained a substantial challenge. 
\par
In recent years, the introduction of machine learning potentials (MLP)~\cite{doi:10.1021/jp9105585,Behler2016,doi:10.1021/acs.jpcc.6b10908,P5673,P6102,P6131} has led to a paradigm change in the development of PESs, since flexible machine learning algorithms allow to combine the accuracy of electronic structure methods with the efficiency of simple empirical potentials. In fact, molecule-surface interactions have been among the early applications of  MLPs~\cite{P0316,P0421,P1388,P1820,P3371}, which in the first years still employed a frozen-surface approximation to restrict the complexity of the PES. Starting with the introduction of high-dimensional neural network potentials (HDNNP) in 2007~\cite{Behler2007,Behler2017}, the construction of MLPs for high-dimensional condensed systems has become possible. Nowadays, many different types of MLPs are available~\cite{P2630,P4862,Smith2017,P4644,P5577,P5794}, which allow to explicitly take all degrees of freedom into account even for systems containing thousands of atoms. This methodical progress now enables to study atom- and molecule-surface interactions with full dimensionality including a mobile surface, and numerous applications have been reported to date~\cite{D0CP03462B,P5069,P5763,P5797,P5972}. 
\par
In spite of these advances, so far most studies have addressed the surfaces of rather simple materials like metals~\cite{P5069,doi:10.1126/science.aad4972,Dorenkamp2018_2,Krueger2015}, solid xenon~\cite{doi:10.1021/acs.jpca.1c03433} or graphene~\cite{Jiang2019,D0CP03462B}, i.e., pure elemental systems, while scattering at binary compounds like oxides or thin surface oxides~\cite{Lecroart2021} has been rarely studied to date. Only recently, a first H-atom beam experiment has been reported for the interaction with \aloxp~\cite{Dorenkamp2018}. Its surface termination is well-studied under high vacuum conditions and does not show any surface reconstruction. The material is also often applied, either as catalyst\cite{B507541F} or as supporting material\cite{catal2010068}, increasing the need for a deeper understanding of processes at its interfaces.
\par
Here, we report a combined theoretical and experimental study addressing the interaction of H-atoms with the \aloxp(0001) surface with the aim to unravel the atomistic details of the scattering process. For this purpose we perform H-atom beam experiments employing four different incident conditions to probe the scattering angle distribution as well as the kinetic energy loss of the atoms leaving the surface. Unlike other systems, such as those involving vibrations or steric effects of scattering molecules, in the present work on an atomic beam only the translational energy is involved. We study the effect of this energy for two different incidence angles. For a deeper analysis, in parallel, we carry out large-scale molecular dynamics simulations based on energies and forces obtained from a HDNNP, which after training provides a first-principles quality description of the scattering process. This combination of methods is used for a detailed analysis of the quality of the PES for different initial beam kinetic energies and incidence angles, which allows us to assess the reliability of the employed exchange correlation functional for scattering at different \alox surface sites.

%%%%%%%%%%%%%%%%%%%%%%%%%%%%%%%%%%%%%%%%%%%%%%%%%%%
\section{Methods \label{sec:method} }
%%%%%%%%%%%%%%%%%%%%%%%%%%%%%%%%%%%%%%%%%%%%%%%%%%%

%%%%%%%%%%%%%%%%%%%%%%%%%%%%%%%%%%%%%%%%%%%%%%%%%%%
\subsection{Experimental methods \label{subsec:exp}}
%%%%%%%%%%%%%%%%%%%%%%%%%%%%%%%%%%%%%%%%%%%%%%%%%%%

The hydrogen atom scattering instrument employed in this work is described in detail in Refs. \cite{10.1063/1.5047674,doi:10.1021/acs.jpca.1c00361}. It utilizes a mono-energetic H-atom beam formed by photolysis of a supersonic jet of hydrogen iodide molecules using an excimer laser operating with KrF (\SI{248}{\nano\meter}). The generated H-atoms pass through a skimmer and two apertures separating two differential pumping regions and hit the sample in the ultrahigh vacuum (UHV) scattering chamber. The sample is mounted on a 6-axis manipulator that allows the variation of the polar and azimuthal incidence angles. Scattered H-atoms are detected by Rydberg atom tagging time-of-flight (TOF)~\cite{Schnieder1991}, where two laser pulses excite the H-atoms to a long-lived (n = 34) Rydberg state. The neutral Rydberg atoms travel \SI{250}{\milli\meter}, pass a grounded mesh and are field-ionized and detected with a multichannel plate (MCP) detector. A multichannel scaler records the TOF distributions that are converted to kinetic energy distributions. The detector is rotatable in the plane containing the H-atom beam and the surface normal, providing TOF spectra at various scattering angles. The alumina sample was cleaned by annealing for several hours in an oxygen atmosphere (\SI{10e-6}{\milli\bar}) at \SI{600}{\celsius}. The cleanliness and structure of the surfaces are monitored by Auger electron spectroscopy (AES) and low-energy electron diffraction (LEED). Based on AES and LEED we conclude that the surface has a (1$\times$1) structure with Al termination, Al-O$_3$-Al trilayers~\cite{Chang1968,Ahn1997,Renaud1998,aloxsurface} and a step density of about 2-4\%. Moreover, it is important to note that depending on the stacking of the trilayers there are two possible terminations related by a mirror operation resulting in two possible incident azimuthal angles $\phi_\mathrm{i}$ with respect to the $[10\bar{1}0]$ surface direction of either \SI{0}{\degree} or \SI{180}{\degree} \cite{WALTERS2000L732} that cannot be distinguished with the methods available in experiment, which needs to be considered in the simulations. Further details about the sample characteristics are given in the SI. The inelastic scattering of the H-atoms was performed for initial kinetic energies of 0.99~eV and 1.92~eV and incidence polar angles of  $\theta_\mathrm{i} =$ \SI{40}{\degree} and $\theta_\mathrm{i} =$ \SI{55}{\degree} resulting in four different sets of scattering conditions.

%%%%%%%%%%%%%%%%%%%%%%%%%%%%%%%%%%%%%%%%%%%%%%%%%%%
\subsection{High-dimensional neural network potentials \label{subsec:HDNNP}}
%%%%%%%%%%%%%%%%%%%%%%%%%%%%%%%%%%%%%%%%%%%%%%%%%%%

The full-dimensional potential energy surface of the hydrogen atom scattering process at \alox is represented by a high-dimensional neural network potential. HDNNPs are a frequently used type of machine learning potential introduced by Behler and Parrinello in 2007~\cite{Behler2007}. In second-generation HDNNPs as employed here~\cite{Behler2021}, the total energy of the system is constructed as a sum of atomic energies,
\begin{eqnarray}
E=\sum_{n=1}^{N_\mathrm{atoms}}E_n\ ,  \label{eq:nnetot}   
\end{eqnarray}
that depend on the local atomic environments up to a cutoff radius $R_{\mathrm{c}}$, which has to be chosen large enough to include all relevant atomic interactions. The positions of all neighboring atoms inside the resulting cutoff spheres are described by vectors of atom-centered symmetry functions (ACSF)\cite{Behler2011}, which are invariant with respect to rotation, translation and permutation of the system and in combination with Eq. \ref{eq:nnetot} allow to explicitly include all atomic degrees of freedom. The atomic energies are then computed as outputs of element-specific atomic neural networks as a function of the respective input ACSF vectors. Since only the Cartesian coordinates of all atoms in the system are required for the calculation of the ACSFs, HDNNPs are able to describe the making and breaking of bonds, which is an essential requirement for studying scattering at surfaces.

The weight parameters of the atomic neural networks are obtained in an iterative training process making use of energies and forces obtained from electronic structure calculations\cite{Behler2015,P6548}. Since atomic energies are not quantum mechanical observables, total energies are used as target properties, and the partitioning into atomic energies is done automatically during the training process. The forces, which are needed for training as well as for running molecular dynamics simulations, can be computed as analytic derivatives of the HDNNP energy. 
More details about HDNNPs, the underlying methodology and typical applications can be found in several reviews~\cite{Behler2014,Behler2015,Behler2017,Behler2021}.

%%%%%%%%%%%%%%%%%%%%%%%%%%%%%%%%%%%%%%%%%%%%%%%%%%%
\section{Computational Details \label{sec:details}}
%%%%%%%%%%%%%%%%%%%%%%%%%%%%%%%%%%%%%%%%%%%%%%%%%%%

%%%%%%%%%%%%%%%%%%%%%%%%%%%%%%%%%%%%%%%%%%%%%%%%%%%
\subsection{Density functional theory calculations}
%%%%%%%%%%%%%%%%%%%%%%%%%%%%%%%%%%%%%%%%%%%%%%%%%%%

The DFT reference energies and forces for training the HDNNP have been calculated using the Fritz-Haber-Institute ab initio molecular simulations (FHI-aims) code (version 171221\_1)\cite{Blum2009} applying the "light" settings for the integration grid and for the basis sets, which consist of numerical atomic orbitals. Spin-polarized calculations were performed for structures containing hydrogen atoms, while \alox bulk and slab structures without hydrogen atoms have been treated as closed-shell systems in a numerically consistent way. $\Gamma$-centred $\mathbf{k}$-point grids have been used for all periodic systems. For the bulk \alox structures containing 40 atoms a 6$\times$6$\times$2 $\mathbf{k}$-point grid was used, while for the (2$\times$2) slab supercells a 2$\times$2$\times$1 $\mathbf{k}$-point grid has been employed. Using these $\mathbf{k}$-point grids, the formation energy of bulk \alox and the adsorption energies of hydrogen atoms at different surface sites are converged to about \SI{0.5}{\milli\electronvolt}. The convergence criteria for the electronic self-consistency of the single point calculations have been set to $\SI{e-6}{\electronvolt}$ for total potential energies and $\SI{e-4}{\electronvolt\per\angstrom}$ for the forces.

Several exchange correlation functionals have been tested in combination with Tkatchenko-Scheffler dispersion corrections~\cite{TS_corr}, including the GGA functionals PBE~\cite{Perdew1996} and RPBE~\cite{RPBE} as well as the PBE0~\cite{P1666} hybrid functional. Figure S3 in the SI shows two one-dimensional cuts through the potential energy surface for a hydrogen atom approaching the surface on top of an aluminium and an oxygen surface site, respectively. For both sites the the observed interaction energies at the potential minimum are within a range of 0.1~eV for all functionals, which is the expected uncertainty with respect to the description of exchange and correlation. This is about one order of magnitude smaller than the lower initial kinetic energy of 0.99~eV used in experiment, even if a  potential energy amount of 0.1~eV would be converted entirely to H-atom kinetic energy. At much lower experimental kinetic energies, however, the accuracy of the employed functional would become highly relevant. At both surface sites the PBE functional is more attractive than RPBE with the minimum region showing no clear trend for a better agreement with the hybrid functional. 
Hence, in the present work, which strongly depends on the description of the repulsive walls due to the high kinetic energies of the H-atoms, we will employ the RPBE functional that is in reasonable agreement with PBE0 for smaller, i.e., repulsive, atom-surface distances. Performing a very large number of PBE0 calculations would be very demanding for the construction of the training set while on the other hand using this hybrid functional does not yield a qualitatively different description of the PES.

%%%%%%%%%%%%%%%%%%%%%%%%%%%%%%%%%%%%%%%%%%%%%%%%%%%
\subsection{Construction of the reference set \label{sec:referenceset}}
%%%%%%%%%%%%%%%%%%%%%%%%%%%%%%%%%%%%%%%%%%%%%%%%%%%

\begin{figure*}[!h]
    \centering
    \includegraphics{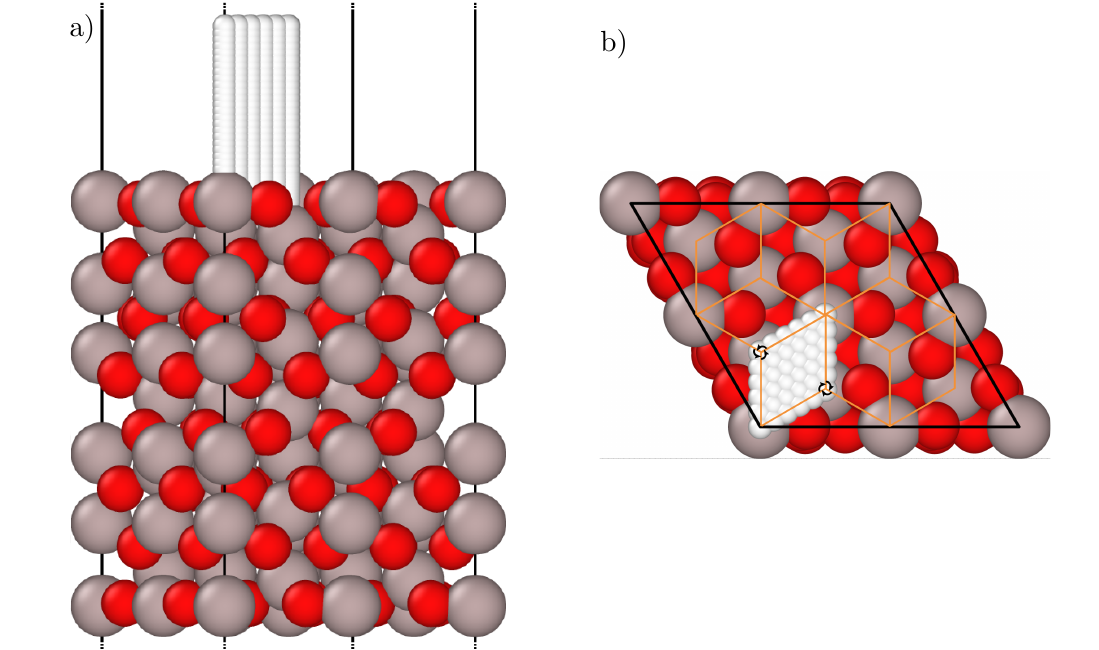}
    \caption{Side (a) and top view (b) of the relaxed (2$\times$2) $\alpha$-Al$_2$O$_3$(0001) supercell slab. Al atoms are shown in grey, O atoms are displayed in red. The white atoms represent the regular H-atom grid employed for the initial screening of the potential-energy surface using a frozen surface. The orange lines in (b) represent the boundaries of the symmetry-unique wedges, and two examples for the three-fold rotational symmetry axes aligned perpendicular to the surface are highlighted. The figures have been created using Ovito version 3.8.4\cite{ovito}.}
    \label{fig:slab_sideways}
\end{figure*}

The reference data set contains bulk geometries and slab structures of the \aloxp(0001) surface, both with and without an H-atom at various positions above the slab. The bulk structures contain one unit cell of \aloxp, while for the slabs (2$\times$2) supercells of the \aloxp(0001) surface unit cell have been constructed with the vacuum aligned in $z$ direction to avoid lateral interactions between the periodic images of the H-atoms. The slabs consist of 8 Al-O$_3$-Al trilayers as shown in Figure \ref{fig:slab_sideways}. The surfaces are terminated by one Al-atom per unit cell on top of an oxygen layer, which is called the ``half metal'' termination and corresponds to the surface structure present in experiment~\cite{Ahn1997,Renaud1998,aloxsurface}. The atoms of the bottom four trilayers have been frozen at their bulk equilibrium positions, while the atoms in the top four layers have been relaxed and are also mobile in the MD simulations, which ensures that the frozen surface atoms in the deep layers are outside the cutoff spheres of the ACSFs describing the hydrogen atom environment. Utilizing an analysis of the atomic interactions based on the Hessian matrix of the slab~\cite{P6312}, we found that the remaining interactions between the deep frozen layers and the H-atom are negligibly small even for H-atom positions close to the surface (see supporting information). The vacuum between the slabs was set to \SI{21}{\angstrom}. At a distance from the surface of about \SI{6.3}{\angstrom} corresponding to the ACSF cutoff only very small forces of roughly \SI{0.003}{\electronvolt\per\angstrom} act on the hydrogen atoms such that atomic interactions at larger separations can be neglected to a good approximation. The functional form of the cutoff function~\cite{Behler2011} ensures a smooth decay of the forces to zero at the cutoff radius in the HDNNP.
The size of the vacuum prevents any notable interactions between the H-atom and the opposite side of the slab in the periodic DFT calculations and the HDNNP even for the largest H-atom-surface distance of 6.8~\AA{} we studied. 

The reference data points were generated combining various approaches. Initially, 133 bulk and 200 clean slab structures were generated with random atomic displacements of up to \SI{0.2}{\angstrom}. Initial HDNNPs based on these structures were used in MD simulations in the canonical ensemble at increasing temperatures up to 400 K to identify important missing surface structures through active learning~\cite{Artrith2012,P5399,P4939}.  
Structures showing an energy or force variance larger than \SI{2.8}{\milli\electronvolt\per\atom} or \SI{0.5}{\electronvolt\per\bohr} (with $a_0$ being the Bohr radius) for different preliminary HDNNPs, respectively, were added to the data set. 

Hydrogen atom positions above the surface were initially included using an equidistant grid with spacing of \SI{0.5}{\angstrom} in the $x$ and $y$ Cartesian directions, starting at the vertical position of the topmost oxygen atom layer and up to \SI{6.75}{\angstrom} above this layer within the symmetry-unique part of the unit cell (s. Fig.~\ref{fig:slab_sideways}) employing the fixed structure of the clean relaxed surface. In the $z$ direction the distances between the H-atom positions are increasing with distance from the surfaces due to the reduced variance in the atomic interactions at larger distances. Up to a distance of \SI{1.0}{\angstrom} from the topmost oxygen layer a vertical grid spacing of \SI{0.125}{\angstrom} has been employed, followed by a vertical spacing of \SI{0.25}{\angstrom} for distances up to the limit of \SI{6.75}{\angstrom}. From this grid, all strongly repulsive structures with a distance between the H-atom and any surface atom below \SI{0.5}{\angstrom} have been discarded.
In the next step, active learning has been performed employing molecular dynamics simulations of the scattering process using the experimental incidence conditions including a mobile surface to sample the surface degrees of freedom. Additional structures obtained in these trajectories were additionally selected by farthest point sampling \cite{farthest_point} as discussed in the SI to improve the representation of rarely visited geometries. 

%%%%%%%%%%%%%%%%%%%%%%%%%%%%%%%%%%%%%%%%%%%%%%%%%%%
\subsection{Construction of the high-dimensional neural network potential}\label{sec:construction}
%%%%%%%%%%%%%%%%%%%%%%%%%%%%%%%%%%%%%%%%%%%%%%%%%%%

For the construction of the HDNNP the RuNNer code (version from August 22, 2019) was used \cite{Behler2015,Behler2017}. The atomic neural networks consist of two hidden layers containing 19 neurons each with $n^\alpha_G$ input neurons for atoms of element $\alpha$ corresponding to the respective number of ACSFs, and one output node providing the atomic energy. The ACSF cutoff radius is $R_c = \SI{12}{\bohr}$ ($\SI{6.35}{\angstrom}$), which is sufficient to include a major part of the atomic interactions as discussed in Section~\ref{sec:referenceset}. A list of the employed atom-centered symmetry functions can be found in the supplementary information. Since there is only one H-atom per structure, ACSFs for the description of H-H interactions have not been  included. For the construction of the HDNNP the atomic energies were removed from the total energies before training such that numerically more favorable binding energies are learned. The HDNNP was then trained to reproduce these binding energies and the DFT atomic force components of the reference structures. The reference data set was randomly split into a training set consisting of about 90~\% of the structures used for adjusting the weight parameters and a testing set of the remaining 10 \% of the structures for assessing the quality of the HDNNP for unknown structures. The parameters of the Kalman filter~\cite{P1308} were set to $\lambda = 0.98000$ and $\nu = 0.99870$. Overall, the structures cover an energy range of \SI{0.446}{\electronvolt ~atom^{-1}}. Atoms with forces up to \SI{13.6}{\electronvolt \angstrom^{-1}} have been included in the data set for monitoring the training errors, but only forces up to \SI{3}{\electronvolt \angstrom^{-1}} have been used to update the weights in the training process.

%%%%%%%%%%%%%%%%%%%%%%%%%%%%%%%%%%%%%%%%%%%%%%%%%%%
\subsection{Molecular dynamics simulations \label{sec:md}}
%%%%%%%%%%%%%%%%%%%%%%%%%%%%%%%%%%%%%%%%%%%%%%%%%%%

To run MD simulations the Large-scale Atomic/Molecular Massively Parallel Simulator (LAMMPS) (version from 16$^\text{th}$ March 2018)\cite{Plimpton1995} including the n2p2 extension for HDNNPs\cite{n2p2} was used. MD simulations for active learning were run in the canonical $NVT$ (bulk and clean surfaces) and in the microcanonical $NVE$ (trajectories for H-atom scattering) ensemble with a time step of $\delta t=\SI{0.5}{\femto\second}$ for simulations without a hydrogen atom and $\delta t=\SI{0.25}{\femto\second}$ for simulations including a hydrogen atom. The velocity Verlet algorithm was used as integrator\cite{Swope1982}. To simulate the $NVT$ ensemble a Nos\'{e}-Hoover thermostat\cite{nose_hoover} with a temperature damping parameter of \SI{0.01}{\pico\second} was used. 
\par
For the MD trajectories of H-atom scattering, i.e., slab structures at room temperature, atomic velocities of the surface atoms were taken from $NVT$ trajectories at \SI{300}{\kelvin} in intervals of \SI{0.1}{\pico\second}. Then, the H-atom was placed at a random lateral position \SI{6.3}{\angstrom} above the surface with velocity vectors corresponding to the experimental kinetic energy, polar and azimuthal angle. The trajectories were run in the $NVE$ ensemble and terminated after \SI{0.5}{\pico\second} or before, when the H-atom surface distance exceeded a value of \SI{7.8}{\angstrom}. Trajectories were considered in the analysis of the kinetic energy loss and the angular distributions if the H-atom was at least at a distance of \SI{7.3}{\angstrom} from the slab at the end of the trajectory. The resulting kinetic energies and scattering angles of the H-atoms were calculated using the velocity vector of the last time step of the trajectory.  The detector counting the H-atom flux is defined as a vector with the specified polar angle $\theta_\mathrm{s}$ towards the surface normal and in plane with the incident azimuthal angle $\phi_\mathrm{i}$ to mimic the position of the detector in the experiment. The experimental detection angle $\gamma$, which is the angle between the velocity vector of the H-atom and the detector, is about \SI{1.5}{\degree}, while in the MD simulations larger angles up to \SI{5}{\degree} have been included to improve statistics, which has very little effect on the obtained results (for a discussion see SI). 

%%%%%%%%%%%%%%%%%%%%%%%%%%%%%%%%%%%%%%%%%%%%%%%%%%%
\subsection{Phonon Calculations}
%%%%%%%%%%%%%%%%%%%%%%%%%%%%%%%%%%%%%%%%%%%%%%%%%%%

Phonon band structures were calculated with the phonopy code (version 2.17.1)\cite{phonopy1,phonopy2} using both, force constants from FHI-aims RPBE DFT calculations as well as force constants calculated using the HDNNP. The force constants were determined using a geometry optimized primitive \alox cell containing 10 atoms. The geometry optimization using the HDNNP  was carried out using LAMMPS~\cite{Plimpton1995} with the n2p2~\cite{n2p2} extension. To determine the correct symmetry, the symmetry precision threshold was set to $5 \cdot 10^{-4}$. The interface between phonopy and RuNNer was created using a Python script.

%%%%%%%%%%%%%%%%%%%%%%%%%%%%%%%%%%%%%%%%%%%%%%%%%%%
\section{Results and Discussion \label{sec:results}}
%%%%%%%%%%%%%%%%%%%%%%%%%%%%%%%%%%%%%%%%%%%%%%%%%%%

%%%%%%%%%%%%%%%%%%%%%%%%%%%%%%%%%%%%%%%%%%%%%%%%%%%
\subsection{Validation of the HDNNP}
%%%%%%%%%%%%%%%%%%%%%%%%%%%%%%%%%%%%%%%%%%%%%%%%%%%

The final data set obtained after several active learning cycles consists of 15,812 structures, which include 808 bulk configurations and 15,004 slab geometries (12,704 with H-atom and 2,300 clean surface structures). The final HDNNP has a root mean square error (RMSE) of \SI{0.746}{\milli\electronvolt ~atom^{-1}} for the energy and \SI{0.103}{\electronvolt\per\angstrom} for the atomic force components for the test data set, which is in the typical order of magnitude of state-of-the-art MLPs. For the training set the RMSE values are \SI{0.257}{\milli\electronvolt ~atom^{-1}} and \SI{0.111}{\electronvolt\per\angstrom}, respectively. Therefore, the HDNNP errors with respect to the reference method are significantly smaller than the uncertainty with respect to the choice of the exchange correlation functional. 
More detailed information about the energy and force errors can be found in the SI. 

\begin{figure}[H]
    \centering
    \includegraphics[width=0.5\textwidth]{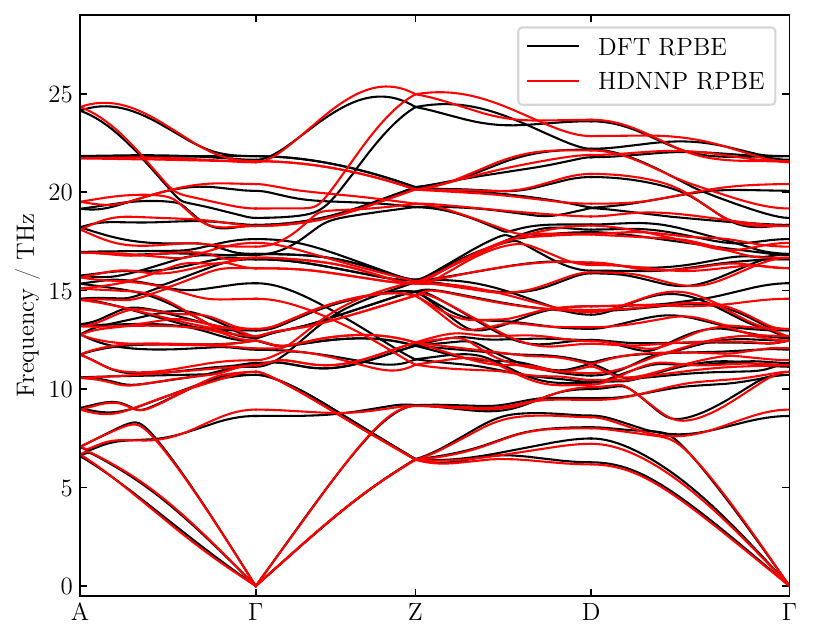}
    \caption{Phonon band structure of bulk \alox computed for the HDNNP and DFT employing the RPBE functional using the phonopy program~\cite{phonopy1,phonopy2}.}
    \label{fig:phonon}
\end{figure}

As monitoring the RMSEs of the energies and forces alone is insufficient to judge on the quality of a potential, in a first validation step we calculated the lattice constants and bulk modulus $B_0$ for bulk \aloxp. As can be seen in Table \ref{tab:lattice}, the HDNNP predictions are in excellent agreement with the underlying RPBE-DFT data, while compared to experiment the RPBE functional slightly overestimates the lattice constants and underestimates the bulk modulus. In a next step, we have computed the phonon band structure of bulk \alox using the phonopy program\cite{phonopy1,phonopy2}. The phonon spectrum obtained from the HDNNP in Figure \ref{fig:phonon} shows excellent agreement with DFT in particular at low frequencies. Overall, the phonon band structure, which is important for energy dissipation after the scattering process, is well represented, and we conclude that the HDNNP provides a reliable description of bulk \aloxp.

\begin{table}
	\centering
	\caption{Calculated and experimental lattice constants $a$ and $c$ and bulk modulus $B_0$ of bulk \alox\cite{Wang}.}
	\begin{tabular}{cccc}
		\toprule
		 Parameter & RPBE & HDNNP & Exp. \\
		\midrule
		  $a$/\AA{} & 4.827 & 4.824 &  4.7554\cite{Wang} \\
		 $c$/\AA{} & 13.145 & 13.139 & 12.991\cite{Wang} \\
		 $B_0$/GPa & 222 & 216 & 253\cite{Anderson} \\
		\bottomrule
	\end{tabular}%
	\label{tab:lattice}%
\end{table}

In order to ascertain the quality of the structural description of the clean surface, we compared the relaxed interlayer distances of the clean surface optimized with RPBE DFT and with the HDNNP. When optimizing the surface with DFT, the topmost three interlayer distances relax from the initial bulk separations of \SI{1.59}{\angstrom} (Al-O$_3$), \SI{1.59}{\angstrom} (O$_3$-Al) and \SI{0.96}{\angstrom} (Al-Al), respectively, to \SI{0.18}{\angstrom} (Al-O$_3$), \SI{1.69}{\angstrom} (O$_3$-Al) and \SI{0.52}{\angstrom} (Al-Al). Consequently, there is a substantial inwards relaxation in particular of the top Al layer, and in the HDNNP relaxation this Al-O$_3$ layer distance is about \SI{0.15}{\angstrom} with an error of only about \SI{0.016}{\angstrom}. For all other interlayer distances the differences between the HDNNP and RPBE results are below \SI{0.01}{\angstrom}.

\begin{figure*}[!h]
    \centering
    \includegraphics{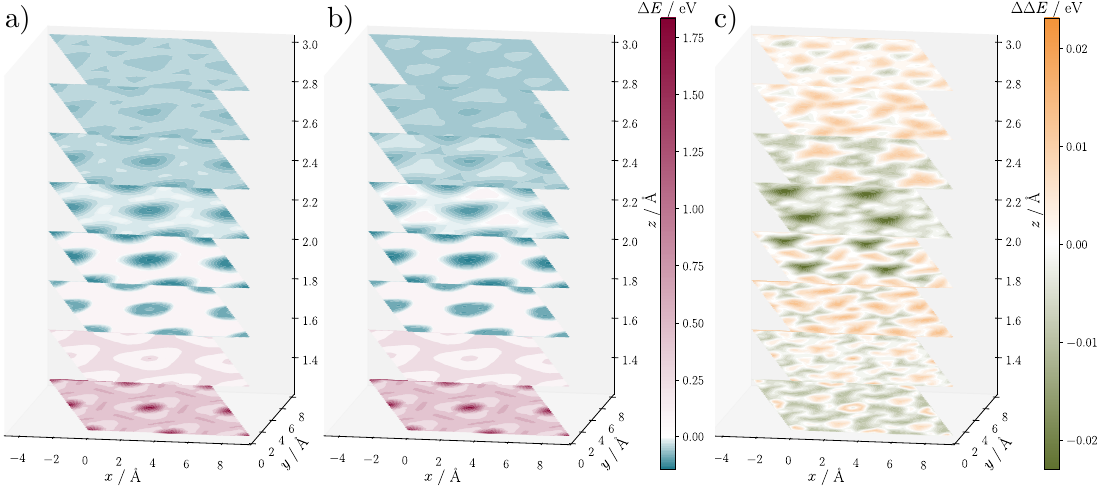}
    \caption{2D cuts at different distances through the PES of the H-atom above a frozen \aloxp(0001) surface. Panel (a) depicts the (2$\times$2) supercell of the RPBE PES while (b) shows the HDNNP PES. Given are relative potential energies $\Delta E$ of the entire system with respect to the energy of a H-atom at infinite separation from the surface. The underlying data points correspond to the regular grid of H-atom positions shown in Fig.~\ref{fig:slab_sideways}.
    Panel (c) shows the difference $\Delta\Delta E$ between RPBE and the HDNNP. Note that in contrast to the energy RMSEs these energies are not normalized per atom in the system.}
    \label{fig:2D_plot_paper}
\end{figure*}

As an initial assessment of the quality of the hydrogen atom-surface interactions, we have analyzed the 3D-PES of a H-atom at different positions above the surface, which, as described in Section~\ref{sec:referenceset}, also formed the starting point of the reference data generation beyond this grid. Figure \ref{fig:2D_plot_paper} shows several 2D cuts through this PES at different H distances from the surface, which is frozen in its relaxed geometry for this purpose. Overall, the RPBE and the HDNNP PESs displayed in Figures~\ref{fig:2D_plot_paper}a and \ref{fig:2D_plot_paper}b are very similar, as can also be seen in the energy difference plot shown in panel (c). The total energy differences are found to be typically between \SI{-0.01}{\electronvolt} and \SI{0.01}{\electronvolt}, which corresponds to an error of \SI{0.06}{\milli\electronvolt \per \atom}. The largest difference is about \SI{-0.04}{\electronvolt} in the cuts at \SI{2.0}{\angstrom} and \SI{2.25}{\angstrom}, i.e., \SI{0.25}{\milli\electronvolt \per \atom}, which is still well below the RMSE of the HDNNP. We conclude that the differences are very small. However, the 3D-PES only probes the interaction between the H-atom and the frozen surface, which is insufficient to check the reliability in MD simulations of the scattering process with fully mobile surface atoms. 
\par
To validate the performance of the HDNNP for full-dimensional trajectories, ab initio MD simulations have been carried out and compared to the HDNNP. The starting structures for the ab inito MD simulations and the atomic velocities for the slab atoms were taken from $NVT$ simulations at \SI{300}{\kelvin} of the clean surface as determined using the HDNNP. The magnitude and direction of the velocities of the H-atoms correspond to the experimental conditions, and we have chosen such conditions that are estimated to result in experimentally detectable scattering events based on preliminary HDNNP trajectories. In total, ten different ab initio MD trajectories have been computed, and the results are shown in Table \ref{tab:ab-initio}. Since even smallest numerical differences result in diverging MD trajectories, a direct comparison of independently computed ab initio and HDNNP trajectories using the same initial conditions is not possible. Thus, for assessing the reliability of the HDNNP, we have recomputed the structures visited in the ab initio trajectories by single point HDNNP calculations and computed the RMSEs of the energies and the $z$-components of the forces acting on the H-atom (s. Table~\ref{tab:ab-initio}). Overall, we find a very good agreement between DFT and the HDNNP for all trajectories with deviations similar to or below the RMSEs of the training and test data sets.

\begin{table}[htb!]
  \centering
  \caption{Initial conditions of the ab initio MD simulations used to validate the HDNNP, including the incident polar angle $\theta_\mathrm{i}$, the incident azimuth angle $\phi_\mathrm{i}$ and incident kinetic energy $E_\mathrm{kin,i}$, as well as the polar angle $\theta_\mathrm{s}$ and the kinetic energy of the scattered atoms $E_\mathrm{kin,s}$. Moreover, the RMSE values of the total energy and of the $z$-component of the H-atom force vector have been computed by comparing the ab initio MD trajectory with subsequent HDNNP single point calculations along the trajectories to quantify the deviations between DFT and the HDNNP. The surface has been equilibrated at \SI{300}{\kelvin}.}
  \label{tab:ab-initio}
  \begin{tabularx}{0.49\textwidth}{XXXXXXXX}
    Name &$\theta_\mathrm{i}$ / \si{\degree} & $\phi_\mathrm{i}$ / \si{\degree} & $E_\mathrm{kin,i}$ / \si{\electronvolt} & $\theta_\mathrm{s}$ / \si{\degree} & $E_\mathrm{kin,s}$ / \si{\electronvolt} & RMSE $E$ / meV atom$^{-1}$ & RMSE $F_\mathrm{H,z}$ / \si{\electronvolt\per\angstrom} \\
    \hline
    MD1 & 40 & 0 & 0.99 & 54 & 0.56 & 0.39 & 0.039\\ 
    MD2 & 40 & 0 & 1.92 & 25 & 1.65 & 0.35 & 0.041\\
    MD3 & 40 & 0 & 1.92 & 44 & 1.69 & 0.32 & 0.037\\
    MD4 & 40 & 0 & 1.92 & 50 & 1.78 & 0.38 & 0.045\\
    MD5 & 40 & 180 & 1.92 & -10 & 1.00 & 0.34 & 0.125\\
    MD6 & 40 & 180 & 1.92 & 32 & 1.00 & 0.33 & 0.047\\
    MD7 & 40 & 180 & 1.92 & 61 & 0.83 & 0.35 & 0.102\\
    MD8 & 40 & 180 & 1.92 & 13 & 0.57 & 0.47 & 0.108 \\
    MD9 & 55 & 0 & 0.99 & 49 & 0.83 & 0.36 & 0.022\\
    MD10 & 55 & 180 & 1.92 & 48 & 1.70 & 0.37 & 0.047\\
    \hline
  \end{tabularx}
\end{table}

\begin{figure*}[htb!]
\centering
\includegraphics{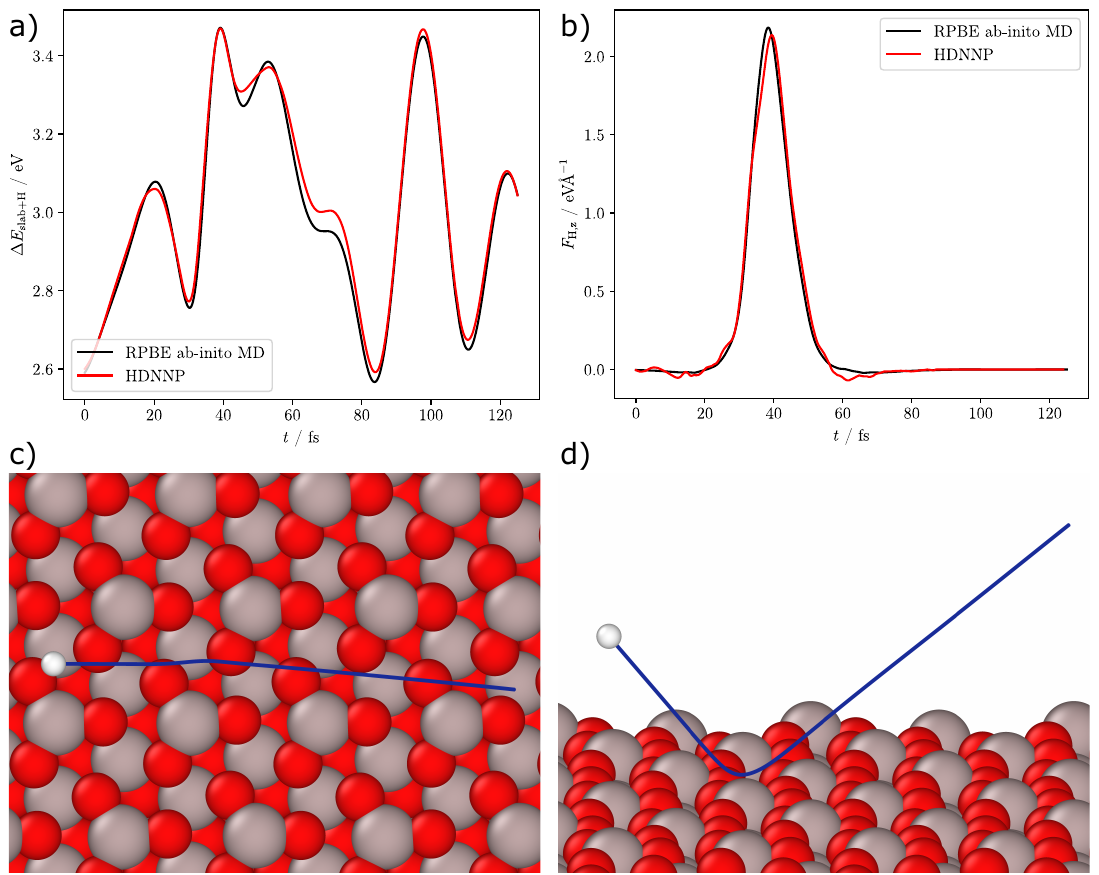}
\caption{Analysis of ab initio MD trajectory MD4 (see Table~\ref{tab:ab-initio}). Panel \textbf{(a)} shows the DFT and HDNNP potential energies of the structures $\Delta E_\mathrm{slab+H}$ with respect to the relaxed slab and the H-atom at infinite distance from the surface vs. the time $t$ of the trajectory. The trajectory was calculated using RPBE DFT and afterwards the structures were recalculated using the HDNNP showing very close agreement. Panel \textbf{(b)} shows the $z$-component of the force vector acting on the H-atom. Panels \textbf{(c)} and \textbf{(d)} show the path (blue) of the H-atom (white) along the surface (top and side views). The surface structure corresponds to the first frame of the simulation, the highlighted H-atom denotes the starting position above the surface.}
\label{fig:ab-inito1}
\end{figure*}

Figure \ref{fig:ab-inito1} shows a detailed analysis of the scattering process for the example of trajectory MD4 with an incident kinetic energy of \SI{1.92}{\electronvolt} and an incident polar angle of \SI{40}{\degree}.
The trajectory was calculated using RPBE DFT and afterwards the structures were recalculated using the HDNNP. A comparison of the DFT and HDNNP potential energies of the system along the trajectory is shown in Figure \ref{fig:ab-inito1}a, while Figure \ref{fig:ab-inito1}b displays a comparison of the DFT and HDNNP $z$-components of the forces acting on the H-atom. Both, energies and forces, are very well represented by the HDNNP along the entire trajectory, which is shown in a top and a side view in panels c) and d). 
We note that the energy deviations are small in comparison to the initial kinetic energy of the collision of 1.92 eV and thus do not significantly affect the scattering process. A similar agreement has been found also for the other trajectories in Table~\ref{tab:ab-initio}.

%%%%%%%%%%%%%%%%%%%%%%%%%%%%%%%%%%%%%%%%%%%%%%%%%%%
\subsection{Comparison of experimental scattering and simulations}
%%%%%%%%%%%%%%%%%%%%%%%%%%%%%%%%%%%%%%%%%%%%%%%%%%%

Having validated the accurate description of the DFT-PES by the HDNNP, we now perform large-scale MD simulations and compare the scattering properties of the \aloxp(0001) surface with experiment using all four scattering conditions as described in Section~\ref{subsec:exp}.  
Due to the close agreement between DFT and the HDNNP, a comparison between experiment and simulation will allow to assess the performance of the employed DFT description of the scattering process.
\par
Concerning the experimental side, in view of the rather high complexity of the \aloxp(0001) surface in comparison to previous atomic beam studies at surfaces of pure elements, we took great care when characterizing the surface prior to the scattering experiments in detail (see SI). Still, we cannot exclude that differences between experiment and simulation may be related to experimental shortcomings, and a comparison of experimental and theoretical results needs to be done with care. 
\par
In the experiment, incidence polar angle, incidence kinetic energy and surface temperature are well-controlled. Also the incident azimuthal angle is experimentally defined, but here we have to consider that the surface can be terminated in two ways as discussed in Section~\ref{subsec:exp}. However, both represent very different surface morphologies for a tilted incidence angle of the beam and thus produce significantly different scattering spectra. Due to the lack of more detailed information, for the comparison between experiment and MD we have computed trajectories of both terminations assuming a 1:1 ratio in experiment, which in fact also provides the best overall agreement. A further discussion of this aspect can be found in the SI.

\begin{figure*}[!h]
    \centering
    \includegraphics{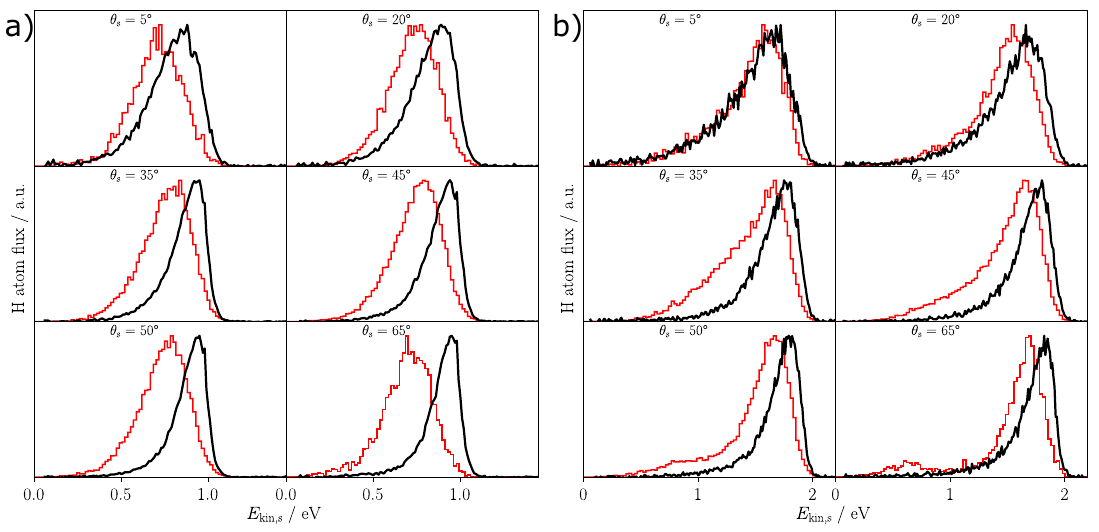}
    \caption{Kinetic energy loss distributions for H-atom scattering at a) $E_\mathrm{kin,i} = $\SI{0.99}{\electronvolt} and $\theta_\mathrm{i} =$ \SI{40}{\degree}, b) $E_\mathrm{kin,i} = $\SI{1.92}{\electronvolt} and $\theta_\mathrm{i} =$ \SI{40}{\degree}. The black line shows the experimentally measured distributions, while the red line shows the theoretical distributions. The detected scattering polar angle $\theta_\mathrm{s}$ is given in each panel. All distributions have been normalized to a maximum of one and the data has been averaged for both azimuthal incident angles of \SI{0}{\degree} and \SI{180}{\degree} in the MD simulations.}
    \label{fig:combo_ekin_loss}
\end{figure*}

Figure \ref{fig:combo_ekin_loss} shows the experimentally obtained kinetic energy loss spectra for the incidence polar angle of \SI{40}{\degree} and both initial kinetic energies $E_\mathrm{kin} = $\SIlist{0.99; 1.92}{\electronvolt} along with the respective data obtained in the MD simulations. The corresponding figure for the incidence polar angle of \SI{55}{\degree} is given in the SI. The distributions have been normalized to a maximum peak height of one. For all incidence conditions a sticking probability of at most 5~\% has been found (s. Table~\ref{tab:sticking}), which therefore does not significantly affect our results. Overall, it can be seen that the simulations have a slight tendency to overestimate the energy loss of the scattered H-atoms with respect to experiment, achieving better agreement for the larger incidence energy. 
\par
\begin{figure*}[!h]
    \centering
    \includegraphics[width=0.95\textwidth]{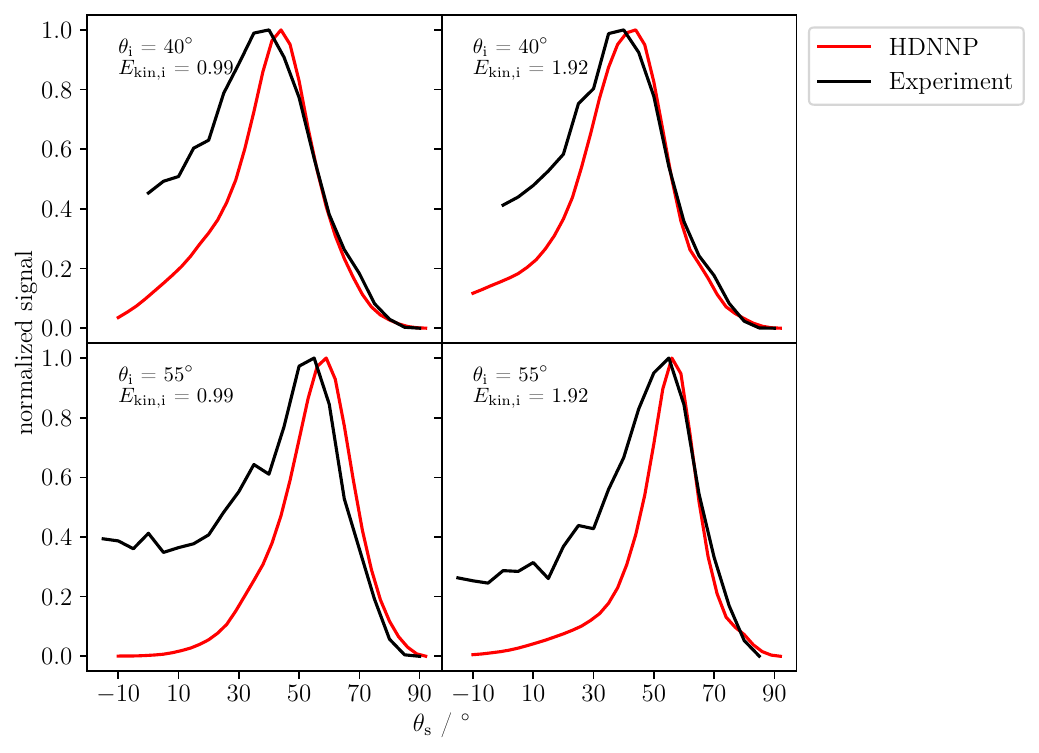}
    \caption{Experimental (black) and computed (red) angular distributions of the scattered H-atoms at different incidence polar angles $\theta_s$ and kinetic energies. The maxima of the distributions have been scaled to one. The data has been averaged for both azimuthal incident angles of \SI{0}{\degree} and \SI{180}{\degree} in the simulations.}
    \label{fig:angular}
\end{figure*}

Figure \ref{fig:angular} compares energy integrated angular distributions of the scattered H-atoms obtained from experiment and simulation. The distributions have been normalized to a maximum of one and the theoretical distributions are again combinations of the two possible surface terminations $\phi_\mathrm{i} = $\SI{0}{\degree} and \SI{180}{\degree}. For all incident conditions the position of the maximum compares well to the experiment and there is very good agreement between experiment and theory for large scattering angles. However, at smaller scattering angles experiment and theory exhibit some deviations, as in experiment scattering  angles close to the surface normal are more frequently found compared to the MD simulations.

\begin{table}[htb!]
  \centering
  \caption{Sticking probabilities $P_{\mathrm{sticking}}$ obtained in the MD simulations for different incidence conditions.}
  \label{tab:sticking}
  \begin{tabularx}{0.3\textwidth}{XXX}
    $E_\mathrm{kin,i}$ / \si{\electronvolt} & $\theta_\mathrm{i}$ / \si{\degree} & $P_\mathrm{sticking}$ \\
    \hline
    0.99 & 40 & 5.0\% \\
    0.99 & 55 & 5.8\% \\
    1.92 & 40 & 4.1\% \\
    1.92 & 55 & 1.3\% \\
    \hline
  \end{tabularx}
\end{table}

%%%%%%%%%%%%%%%%%%%%%%%%%%%%%%%%%%%%%%%%%%%%%%%%%%%
\subsection{Discussion}
%%%%%%%%%%%%%%%%%%%%%%%%%%%%%%%%%%%%%%%%%%%%%%%%%%%

\begin{table}[htb!]
  \centering
  \caption{Relative average energy losses obtained in MD simulations (sim.) and experiment (exp.) compared to the Baule limit (BL) for $\theta_\mathrm{i}=$ \SI{40}{\degree} and $\theta_\mathrm{s}=$ \SI{50}{\degree}.}
  \label{tab:Baule}
  \begin{tabularx}{0.49\textwidth}{XXXXX}
    $E_\mathrm{kin,i}$ / \si{\electronvolt} & sim. & exp. & BL, O-site & BL, Al-site \\
    \hline
    0.99 & 0.26 & 0.12 & 0.22 & 0.11 \\ 
    1.92 & 0.22 & 0.14 & 0.22 & 0.11 \\
    \hline
  \end{tabularx}
\end{table}

Although overall a reasonable agreement has been obtained, there are some deviations between the experimental data and their counterparts determined by molecular dynamics as in the simulations there is a larger kinetic energy loss (Fig.~\ref{fig:combo_ekin_loss}) in particular for the lower incidence kinetic energy as well as a narrower angular distribution of the scattered atoms (Fig.~\ref{fig:angular}) compared to the atomic beam results. However, in previous work, good agreement between experiment and simulation was achieved for well-defined, single-element surfaces, i.e., solid Xe\cite{doi:10.1021/acs.jpca.1c03433}, graphene\cite{D0CP03462B} and several fcc metal surfaces~\cite{Dorenkamp2018_2}. The more pronounced discrepancies in the present study may have several reasons.
\par
First, it has to be ensured that the surface structure used in the simulations is a good representation of the surface that is present in experiment. In case of the \aloxp(0001) surface, controlling the surface structure is more challenging, as, e.g., there are two possible terminations, which are geometrically inequivalent for the scattering atoms and different ratios of both terminations is essentially unknown. The effect of this ratio on the simulation outcome is discussed in the SI. Moreover, the effect of steps and other surface imperfections is not included in the simulations.
\par
Concerning the simulations, the accuracy of the reference electronic structure method poses a limit for the quality of the HDNNP that can be achieved. In the present work a GGA functional has been employed as our initial investigations have shown that for the investigated geometries there are no fundamental differences between the quality of a GGA description and the PBE0 functional in view of the high atomic kinetic energies. However, due to the substantial computational costs of the hybrid functional it has not been possible to construct a full-dimensional HDNNP based on PBE0, and consequently parts of the PES exhibiting larger deviations might have been missed.

We will start discussing the results of the simulations before we address the deviations between experiment and theory in more detail. 
A first step to gain some basic information about the scattering mechanism is to compare the average energy loss with the predictions of the simple Baule model. The Baule limit describes the maximum energy loss that can be expected in a single-atom surface collision. It is based on a zero impact parameter collision of the impinging H-atom with a resting surface atom. An energy loss significantly larger than the Baule limit is a clear indication for a complex scattering mechanism, for example multiple collisions or energy transfer to several atoms in a single collision. Table \ref{tab:Baule} compares the Baule limit for the oxygen and the aluminum site to the average energy losses observed in experiment and theory. It can clearly be seen that the simulation, especially for an incidence kinetic energy of 0.99\,eV, yields a larger relative energy loss than the simple Baule limit, while in experiment a loss below the Baule limit is observed. This observation indicates that there might be a complex scattering mechanism in the simulations that is not necessarily expected based on the experimental data. 

\begin{figure}[ht]
    \centering
    \includegraphics[width=0.5\textwidth]{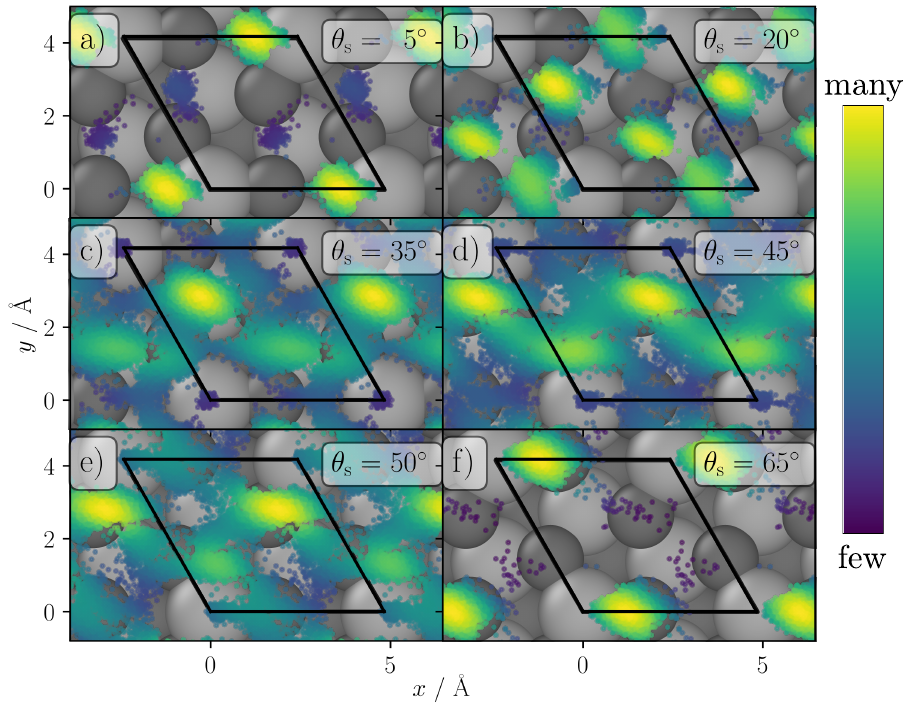}
    \caption{Impact positions of the H-atoms at the surface for the trajectories with incidence conditions $\theta_i = 40^\circ$, $\phi_\mathrm{i} = $\SI{0}{\degree} and an initial kinetic energy of \SI{1.92}{\electronvolt}. For reference the equilibrated surface is shown. The black lines highlight the surface unit cell, darker grey circles are oxygen atoms, lighter grey circles are Al atoms. The brighter the color of the points, the higher is the density of impact sites.}
    \label{fig:xy_lowest}
\end{figure}

\begin{figure}[ht]
    \centering
    \includegraphics[width=0.5\textwidth]{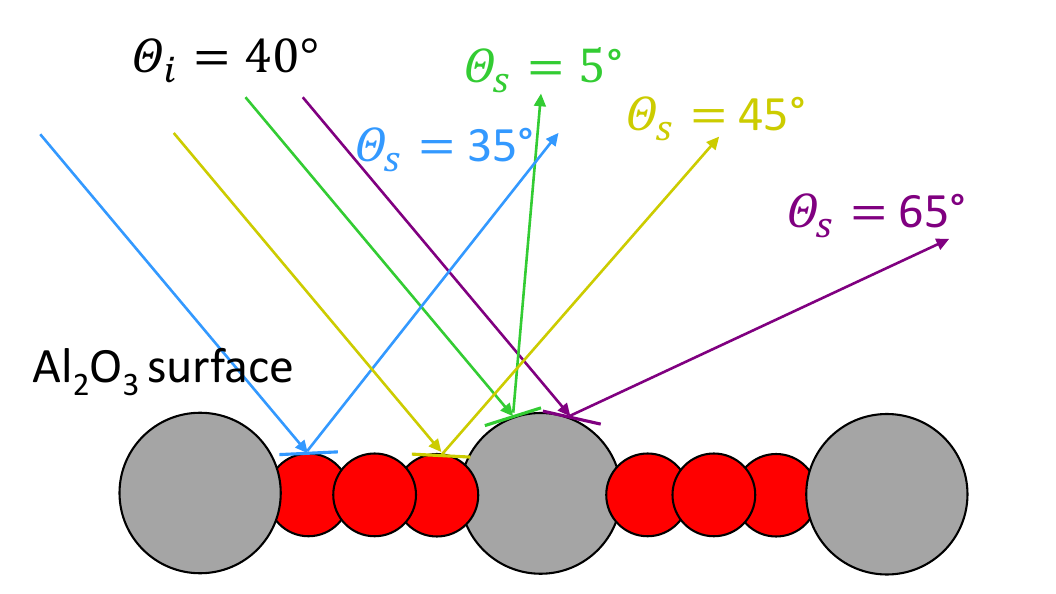}
    \caption{Model of the surface geometry resulting in different scattering angles for elastic scattering in case of an incidence polar angle of $\theta_\mathrm{i} = \SI{40}{\degree}$ .}
    \label{fig:scattering_schematic}
\end{figure}

To understand the reason for the large energy loss observed in simulation we took a closer look at the scattering trajectories leading to high energy losses and analysed them in detail. First, we extracted the geometry of closest approach of the H-atom to the surface for each trajectory to allocate the surface impact sites and to correlated them with the energy loss and scattering angle. In Figure \ref{fig:xy_lowest} the closest point of approach for one incident condition, $\theta_\mathrm{i} = \SI{40}{\degree}$ and $E_\mathrm{kin} = \SI{1.92}{\electronvolt}$, is plotted. The black lines highlight a surface unit cell of \alox. Although the surface is mobile at the surface temperature of \SI{300}{\kelvin}, the estimated location is accurate enough to analyse the origin of the bounce and the trajectories show a clear trend. In trajectories leading to small scattering angles the H-atom is mainly scattering from the topmost Al atoms, at medium angles the scattering predominately takes place at the O atoms and hollow sites between them, while at the largest scattering angles the scattering is again mainly occurring at the Al atoms. This scattering behavior can be explained by rather simple geometric considerations. In Figure \ref{fig:scattering_schematic} a schematic representation of elastic scattering processes at different surface sites is shown for an incidence angle of $\theta_\mathrm{i} = \SI{40}{\degree}$. The top-layer Al atoms effectively shield the O layer preventing large and small scattering angles for the case of scattering at oxygen atoms, while collisions with the Al atoms can result in very low and large scattering angles. The rather flat oxygen layer thus only allows for scattering at angles close to the medium-sized incidence angle. These geometric conditions also apply to the other investigated incident conditions. Best agreement between simulation and experiment for $E_\mathrm{kin} = \SI{1.92}{\electronvolt}$ is achieved for small and large scattering angels for which scattering from Al dominates. For intermediate scattering angles, where scattering from O dominates, the simulation results deviate more strongly from experiment. This might be a hint for a less accurate description of the scattering process at the oxygen sites. Since the comparison of the HDNNP trajectories with ab initio MD does show a similar good agreement for oxygen and aluminium sites, the most likely explanation thus is a less accurate description of the hydrogen-oxygen interaction by the RPBE functional.

Having identified that trajectories initially scattered at oxygen atoms tend to show a too large energy loss, we took a closer look at representative trajectories to get a better understanding of its origin. Specifically, we analyzed the atomic kinetic energies and the changes of the positions of the surface atoms in selected trajectories by comparing them to trajectories with the same initial surface atom velocities but without the presence of a H-atom. In this way, we are able to see how all surface atoms are influenced by the collision with the H-atom. In this analysis we observed that if the H-atom hits an oxygen atom on the surface, the kinetic energy is transferred not only to one oxygen atom but is rapidly distributed over several neighboring surface atoms. Due to the involvement of multiple indirect collision partners during the collision with the O-atoms, the energy loss is increased compared to an idealized single-atom collision. Figure S15 in the SI illustrates how the kinetic energy of the H-atom is efficiently transferred to multiple surface atoms. A similar mechanism has been observed for H-atom scattering from graphene\cite{Jiang2019}. If, however, the H-atom is initially scattered at an Al-atom, the energy is mostly transferred to this Al with an overall much smaller energy loss in a more elastic process. This efficient energy transfer is also seen in \textit{ab-initio} MD simulations supporting that this effect is not a consequence of inaccuracies in the HDNNP. 

One possible source of error contributing to deviations between experiment and simulation are surface imperfections. Although we took great care in experiment to prepare a well-defined surface, some defects are unavoidably present. From AFM measurements we know that the step density of the surface is about 2-4\%. Furthermore, it is known that the \aloxp (0001) surface loses oxygen in an UHV environment. A significant O loss would lead to surface reconstruction and can be excluded, since we observed the (1$\times$1) LEED pattern at all times. Still, oxygen vacancies will be present at the surface. Furthermore, the presence of hydrogen on the surface can not be excluded\cite{Ahn1997} and may affect the energy loss of the scattered hydrogen atoms. The deviation in the angular distributions is a clear indication that the surface is not perfectly flat. Steps and defects increase scattering normal to the surface, exactly what we observe (s. Figure \ref{fig:angular}). However, our finding that the energy loss in experiment is smaller than in simulations can not easily be assigned to surface defects, as surface defects commonly lead to an increased energy loss, for example due to a higher probability of multi-bounce scattering. However, since a complex scattering mechanism involving many atoms is causing the larger energy loss at the oxygen sites in the simulations, it cannot be excluded that surface defects could change or even prohibit such a process.

The second point to be addressed is the general accuracy of the RPBE exchange correlation functional, which determines the interaction strength in between the surface atoms and between the surface atoms and the H atom. The best agreement between the experimental and theoretical kinetic energy distributions is obtained at low scattering angles with high kinetic energies. The largest difference in mean kinetic energy loss is present at large scattering angles and low kinetic incident energies, especially the scattering at $E_\mathrm{kin,i}=\SI{0.99}{\electronvolt}$ and $\theta_\mathrm{s}=\SI{65}{\degree}$. Both cases represent extreme cases of H-atom-surface interaction. When the H-atom scatters at high kinetic energies and low scattering angles, the effective interaction time is minimal. However, when the scattering angle is large and the kinetic energy is low, the interaction time with the surface is significantly longer. As the kinetic energy distributions for high kinetic energies are in better agreement with the experiment compared to the distributions for low kinetic energies, apparently inaccuracies of the reference electronic structure method tend to accumulate with interaction time. 
Moreover, the bulk modulus of \alox obtained in the DFT calculations is about 10\% smaller than the experimental value (Table \ref{tab:lattice}). This softer description of \alox by the RBPE functional may lead to an overestimation of the kinetic energy loss.

%%%%%%%%%%%%%%%%%%%%%%%%%%%%%%%%%%%%%%%%%%%%%%%%%%%%
\section{Summary}
%%%%%%%%%%%%%%%%%%%%%%%%%%%%%%%%%%%%%%%%%%%%%%%%%%%%

We have studied the interaction of hydrogen atoms with the \aloxp(0001) surface using a combination of large-scale molecular dynamics simulations based on a high-dimensional neural network potential trained to DFT data employing the RPBE functional and highly accurate H-atom beam scattering experiments under UHV conditions. Two different initial kinetic energies and two different incidence polar angles have been investigated. 
Best agreement between experiment and theory is found for large initial kinetic energies and very low and high scattering angles, which we find to arise from scattering at top-layer aluminium atoms. Scattering at lower initial kinetic energies results in a larger loss of kinetic energy in the MD trajectories compared to experiment, and in general also scattering at oxygen sites seems to result in larger discrepancies between experiment and theory. As a careful validation of the multidimensional PES representation by the HDNNP showed a very close agreement with the underlying DFT reference calculations, we consider limitations in the accuracy of the employed RPBE functional as the most likely explanation of the observed deviations. Although, in view of the high kinetic energies, the differences between different exchange correlation functionals are small, more complex scattering mechanisms at the oxygen sites might increase the sensitivity of the simulations to subtle differences in the description of the atomic interactions. Moreover, future work should address the possible role of surface imperfections like steps, oxygen vacancies and adsorbed hydrogen atoms, which have not been included in the present work, on the energy loss and scattering mechanisms, which are more complex at the surfaces of binary compounds like \alox compared to frequently studied elemental surfaces.

%%%%%%%%%%%%%%%%%%%%%%%%%%%%%%%%%%%%%%%%%%%%%%%%%%%%
\section{Acknowledgements}
%%%%%%%%%%%%%%%%%%%%%%%%%%%%%%%%%%%%%%%%%%%%%%%%%%%%

O.B. is thankful to Alec M. Wodtke for continuous support in every aspect of the work. We thank Alexander Kandratsenka and Alec M. Wodtke for stimulating discussions. This project has been funded by the Deutsche Forschungsgemeinschaft (DFG, German Research Foundation) under grant numbers 389479699/GRK2455 BENCh and 217133147/SFB 1073, project A04.

%%%END OF MAIN TEXT%%%

\bibliography{bibliography} %You need to replace "rsc" on this line with the name of your .bib file
\bibliographystyle{rsc} %the RSC's .bst file

\end{document}